\documentstyle[preprint,prc,aps,epsfig]{revtex}
\tightenlines
\everymath={\displaystyle}
\begin{document}
\title{On a three-body confinement force in
hadron spectroscopy}

\author{S. Pepin\footnote{e-mail address: pepin@daniel.mpi-hd.mpg.de}}
\address{Max-Planck-Institut f\"ur Kernphysik,
Postfach 103980, D-69029 Heidelberg,
Germany}
\author{Fl. Stancu\footnote{e-mail address: fstancu@ulg.ac.be
}}
\address{University of Li\`ege, Institute of Physics B5, Sart Tilman,
B-4000 Li\`ege 1, Belgium}
\address{ECT*, Strada delle Tabarelle, 286, I-38050 Villazzano,
Trento, Italy}

\date{\today}
\maketitle

\begin{abstract}
\baselineskip=0.50cm
Recently it has been argued that a three-body colour confinement interaction
can affect the stability condition of a three-quark system and
the spectrum of a tetraquark described by any constituent quark model.
Here we discuss the role of a three-body colour confinement
interaction in a simple quark model
and present some of its implications for the spectra of baryons, tetraquarks
and six-quark systems.
\end{abstract}
\section{Introduction}
In the strong coupling limit the SU$_{C}$(3) colour gauge group leads to a
three-body colour confinement interaction in baryons \cite{CKP}.
In SU(3) lattice QCD the static three-quark potential can be measured
with the help of 3-quark Wilson loop operators (see for example
\cite{DOSCH,GROMES,BALI,TAKAHASHI,ALEXANDROU}).
In recent lattice calculations \cite{BALI,TAKAHASHI,ALEXANDROU},
the ground state potential of a three-quark system has been extracted
as a sum of a two-body Coulomb plus a three-body interaction
confinement. These studies lead however to rather contradictory
results; one of them \cite{TAKAHASHI} gives support to the Y-type flux tube
picture of Ref. \cite{CKP}, while the others favour the $\Delta$ ansatz (where
the three-quark potential consists in a sum of two-body components).

The interaction potential obtained in these calculations corresponds
in any case to the colourless ground state only and no information from lattice
QCD about colour octets is available so far.
In practice, for simplicity, in quark models
as e.g. that of \cite{CKP} the confinement is treated
approximately as a two-body colour operator. This can be expressed in
terms of the quadratic (Casimir) invariant operator of SU(3).

This three-body colour confinement interaction should not be confused with
the three-body force \cite{DOROKHOV,OT}
associated to the instanton 't Hooft's interaction,
which in
the nonrelativistic limit contains a colour operator similar
to the one introduced below,
but is of short range, in contradistinction to confinement forces.
The instanton-induced three-body force cancels for three quarks in a colour
singlet state and is moreover only effective
if the three particles are in a flavour singlet state.
Another short-range three-body force has also been introduced in
\cite{Des92}
on a purely phenomenological ground. This force has a simple
scalar structure and accounts for a better description of the
Roper resonance.

Based on the algebraic
argument that SU$_{C}$(3) is an exact symmetry of QCD,
which implies that any quark model Hamiltonian inspired by QCD can be
written in terms of SU(3) invariant operators, a three-quark potential
that depends on the cubic invariant operator of SU(3) has recently
\cite{DMI} been added to the usual two-body confinement.
Its implications on the spectrum of ordinary $q^3$ and exotic $q^2
\overline {q}^2$ hadrons have been considered.
In particular turning on a three-body force with
an appropriate strength denoted by $c$, it was shown that:
1) the three $q^3$ colour states, namely
${\bf 1}$, ${\bf 8}$ and ${\bf 10}$ appear in the correct order,
${\bf 1}$ being the lowest one, as it should be, 2) in the
$q^2 \overline q^2$ system the three-body interaction brings
distinct contributions to the two possible colour singlet states,
by enhancing the binding in one and diminishing it in the other,
depending on the sign of $c$.

In this study we rederive some of the relations found in Ref. \cite{DMI}
and discuss explicitly the role of confining three-body forces
in $q^3$, $q^2 {\overline q}^2$ and $q^6$ systems.
In the simple framework of a harmonic confinement
we show that in the $q^3$ system there is a competition between the
three-body force and the kinetic energy in rising the energy of
the colour octet states.
By introducing singlet-singlet and octet-octet $q \overline q$
coupled pairs we show that the range of
values of the strength of the 3-body force giving a correct
spectrum for a $q^3$ system also
favourably affects the spectrum of a tetraquark
$q^2 \overline q^2$. The results on the role of a 3-body confining force
in a
$q^6$ system, relevant for
the $NN$ problem are entirely new.
\section{The three-body force for baryons}
In this section we recall and discuss
the findings of Ref. \cite{DMI} in relation
with a three-body interaction of type
\begin{equation}\label{3BTOTAL}
V_{3b}=V_{ijk} = {\mathcal V}_{ijk} {\mathcal C}_{ijk}~,
\end{equation}
with
\begin{equation}\label{HO}
 {\mathcal V}_{ijk} = \frac{1}{2}~c~m \omega^2~[(r_i-r_j)^2
+(r_j-r_k)^2+(r_k-r_i)^2]~,
\end{equation}
where $c$ is a strength parameter
and ${\mathcal C}_{ijk}$ a colour operator of type
\begin{equation}\label{3B}
{\mathcal C}_{ijk} = d^{abc}~F^{a}_{i} ~F^{b}_{j}~ F^{c}_{k}~,
\end{equation}
where $F^{a}_{i} = \frac{1}{2} \lambda^{a}_{i} $ is the colour charge
operator of the quark $i$ and $d^{abc}$ some real constants,
symmetric under any permutation of indices and
defined by the anticommutator of the Gell-Mann matrices $\lambda^a$ as
\begin{equation}\label{ANTIC}
\{ \lambda^a,  \lambda^b \} = 2 d^{abc}~\lambda^{c} ~.
\end{equation}
These constants satisfy the following orthogonality relation
\begin{equation}\label{ORTHOG}
d^{abc}~d^{abe} = \frac{5}{3}~ \delta_{ce}
\end{equation}
The operator (\ref{3B}) can be expressed in terms of the two
independent invariant operators of SU(3) as \cite{DMI}
(for a proof see Appendix A)
\begin{equation}\label{EQ4}
d^{abc}~F^{a}_{i} ~F^{b}_{j}~ F^{c}_{k} =
\frac{1}{6}~ [~  C^{(3)}_{i+j+k} - \frac{5}{2} C^{(2)}_{i+j+k} +
\frac{20}{3}~ ]
\end{equation}
where we slightly changed the notation of \cite{DMI},
by writing $C^{(2)}$ instead of $C^{(1)}$ for the quadratic invariant and
$C^{(3)}$ instead of $C^{(2)}$ for the cubic invariant.
For a given irrep of SU(3) labelled by $(\lambda \mu)$,
the eigenvalues of these invariants are
\begin{equation}
\langle C^{(2)} \rangle = \frac{1}{3} (\lambda^2 + \mu^2 + \lambda \mu
+ 3 \lambda + 3 \mu )
\end{equation}
and  (see for example Refs. \cite{ED} or \cite{FS})
\begin{equation}\label{AUTOG}
\langle C^{(3)} \rangle = \frac{1}{18} (\lambda - \mu)
(2 \lambda + \mu + 3)(\lambda + 2 \mu + 3)~.
\end{equation}
Then for a $q^3$ system the expectation values of (\ref{EQ4}) are
$\frac{10}{9},~ -\frac{5}{36},~ \frac{1}{9}$ for a singlet ($\lambda \mu$) =
(00), octet ($\lambda \mu$) = (11) and decuplet ($\lambda \mu$) = (30)
states respectively. These are the coefficients appearing in the
last term of Eq. (\ref{colour}) below.

Turning on a 2-body confining interaction, which ensures stability for
a $q \overline q$ pair and adding
the 3-body confining interaction (\ref{3BTOTAL})-(\ref{3B})
of strength $c$ relative to
the 2-body one, in Ref. \cite{DMI}
it was found that the spectrum of a $q^3$ system is correctly described provided
\begin{equation}\label{INEQ}
-\frac{3}{2} < c < \frac{2}{5}
\end{equation}
The closer $c$ is to the lower limit, the larger is the gap between the
colour octet and singlet states. To see this, let us
consider the Hamiltonian
\begin{equation}\label{HAMILT}
H = T + V_{2b} + V_{3b}
\end{equation}
where T is the kinetic energy and $V_{2b}$ a 2-body confinement
interaction of the form
\begin{equation}\label{V2b}
V_{2b} = \sum\limits_{i<j} V_{ij}~(c_1 + \frac{4}{3} + F^{a}_{i} F^{a}_{j})
\end{equation}
containing an arbitrary constant $c_1$ which we set equal to 1 as
in Ref. \cite{DMI} and take
\begin{equation}
V_{ij} =  \frac{1}{2}~m \omega^2~(r_i-r_j)^2
\end{equation}
$V_{3b}$ is the 3-body confinement interaction
of Eqs. (\ref{3BTOTAL})-(\ref{3B}).
Performing integration in the colour space and expressing $H$ in terms of
the internal
coordinates $\vec{\rho} = (\vec{r_1} - \vec{r_2})/\sqrt{2}$ and
$\vec{\lambda} = (\vec{r_1} + \vec{r_2} - 2 \vec{r_3})/\sqrt{6}$, we have:
\begin{equation}
H = 3 m - \frac{\hbar^2}{2 m} (\nabla_{\rho}^{2} + \nabla_{\lambda}^{2})
+ \frac{3}{2} m \omega^2 \chi_i (\rho^2 + \lambda^2)
\end{equation}
with
\begin{equation}
\chi_i = \left\{ \renewcommand{\arraystretch}{2}
\begin{array}{cl}
 \frac{5}{3} + \frac{10}{9} c  &\hspace{1.5cm} \mbox{ i=1 (singlet)} \\
 \frac{13}{6} - \frac{5}{36} c & \hspace{1.5cm} \mbox{ i=8 (octet)} \\
 \frac{8}{3} + \frac{1}{9} c  & \hspace{1.5cm} \mbox{ i=10 (decuplet)}
\end{array} \right.
\label{colour}
\end{equation}
In the expressions of $\chi_i$ ($i = 1,8$ or 10), the first and second
terms stem from the colour part of $V_{2b}$ and $V_{3b}$ respectively.

We search now for solutions of $H$. In
order to satisfy the Pauli principle, in the lowest
colour-singlet state the
quarks should be in a $s^3$ configuration. This implies that the
orbital wave function is symmetric and one can take:
\begin{equation}
\phi_{00}= \frac{1}{\pi^{3/2} b^3} \exp{[-(\rho^2 + \lambda^2)/(2b^2)]}
\end{equation}
with $b$ a variational parameter.

\noindent The expectation value of $H$ is then:
\begin{equation}
E_1 = 3 m + \frac{3 \hbar^2}{2 m b^2} + \frac{9}{2} m \omega^2 b^2 \chi_1
\label{E1}
\end{equation}
The minimization with respect to $b^2$ gives:
\begin{equation}
b_1^2 = \frac{b_0^2}{\sqrt{3 \chi_1}}
\label{b1}
\end{equation}
with $b_0^2 = \frac{\hbar}{m \omega}$.
The energy of the singlet state is thus:
\begin{equation}
E_1 = 3 m + \frac{3 \hbar^2}{m b_1^2} = 3 m + 3 \hbar \omega \sqrt{3 \chi_1}
\end{equation}

The colour-octet {\bf 8} must be combined with an $s^2p$ configuration in order
to satisfy the Pauli principle. Indeed, the spin-flavour part being
totally symmetric, as for the nucleon ground state, the orbital-colour
part must be antisymmetric, i.e. it is of the form:
\begin{equation}
\Psi_8 = \frac{1}{\sqrt{2}} ( \phi_{10}^\rho C^{\lambda} - \phi_{10}^\lambda
 C^\rho)
\end{equation}
where
\begin{eqnarray}
 \phi_{10}^\rho &=& \left( \frac{2}{\pi^3 b^8} \right)^{1/2} \rho_z
\exp{[- (\rho^2 + \lambda^2)/(2 b^2)]} \\
  \phi_{10}^\lambda &=& \left( \frac{2}{\pi^3 b^8} \right)^{1/2} \lambda_z
\exp{[- (\rho^2 + \lambda^2)/(2 b^2)]}
\end{eqnarray}
are the mixed orbital symmetry states with one unit of angular excitation
and
$C^\rho$, $C^\lambda$ their colour counter-parts. The corresponding
eigenvalue of $H$ is:
\begin{equation}
E_8 = 3 m + \frac{2 \hbar^2}{m b^2} + 6 m \omega^2 b^2 \chi_8
\label{E8}
\end{equation}
and the minimization with respect to $b^2$ gives:
\begin{equation}
b_8^2 = \frac{b_0^2}{\sqrt{3 \chi_8}}
\end{equation}
which leads to
\begin{equation}
E_8 = 3 m + \frac{4 \hbar^2}{m b_8^2} = 3 m + 4 \hbar \omega \sqrt{3 \chi_8}
\end{equation}
Accordingly the gap $\Delta E$ between the octet and singlet states is:
\begin{equation}\label{GAP81}
\Delta E = \hbar \omega ( 4 \sqrt{3 \chi_8} - 3 \sqrt{3 \chi_1})
\end{equation}
The largest value of $\Delta E$ corresponds to $\chi_1 =0$, i.e. to
$c= - 1.5$, in which case $\Delta E \simeq 10.7 \hbar \omega$. For $c = 0$
(no three-body force) we would have $\Delta E \simeq 3.5 \hbar \omega$. This
means that the gap is enlarged by a negative
$c$ and triples for the limiting value $c=-1.5$.

The colour state ${\bf 10}$ requires an antisymmetric orbital state in
order to satisfy the Pauli principle. Its form is (see e.g.  Ref.\cite{FS},
 chap. 10):
\begin{equation}
\phi_{10}^A =(\frac{1}{2 \pi^3 b^{10}})^{1/2} (\rho_+ \lambda_{-}
- \rho_- \lambda_{+}) \exp{[-(\rho^2 + \lambda^2)/(2b^2)]}
\end{equation}
with $\rho_{\pm} = \rho_x \pm i \rho_y$, etc. The subscript 10 means total
L=1, M=0 as above. This is the only value of L allowed by an antisymmetric
state built from the configuration $sp^2$. Proceeding in a similar way as
for the two previous cases, one gets:
\begin{equation}
b_{10}^2 = \frac{b_0^2}{\sqrt{3 \chi_{10}}}
\end{equation}
and hence
\begin{equation}
E_{10} = 3 m + 5 \hbar \omega \sqrt{3 \chi_{10}}
\end{equation}
The gap between the decuplet and the octet state is thus:
\begin{equation}
E_{10} - E_8 = \hbar \omega (5 \sqrt{3 \chi_{10}} - 4 \sqrt{3 \chi_{8}})
\end{equation}
For $c = -1.5$, one has $E_{10} - E_8 \simeq 3 \hbar \omega$, i.e.
this state is located above the octet, as expected, with quite a large gap
for a limiting value of $c$.

    Let us now evaluate the octet-singlet gap (\ref{GAP81}) by
taking c = -1.43 as in \cite{DMI}. There is of course some arbitrariness
in choosing $\hbar \omega$. One can take for example
$b = 0.437$ fm and $m = 0.340$ GeV, as typical values for
a quark model (see e.g. Ref. \cite{SPG}), which give
 $\hbar \omega = $ 0.6 GeV. This implies a gap
$\Delta E \approx $ 5.5 GeV. For $c =$ 0 (no three-body force)
one would have
$\Delta E =  3.5 \hbar \omega \approx $ 2.1 GeV.


Before ending this section we should note that there is some
arbitrariness in fixing the lower limit of the
range of $c$ as given by the inequality (\ref{INEQ}).
This limit is related to the choice of the arbitrary constant $c_1$
which has been set equal to 1 in Eq. (\ref{V2b}). But
taking for example $c_1 = 4/3$,
which is another good choice in constituent quark models, one gets
$\chi_1 = 2 + \frac{10}{9} c$. The stability condition for the singlet would
then gives
\begin{equation}
-\frac{9}{5} < c
\end{equation}
i. e. a different lower limit, slightly more favourable than the one
of inequality (\ref{INEQ}) because with $c =  - 9/5$ and the new expressions
for $\chi_1, \chi_8$ one gets
$\Delta E \simeq 11.5 \hbar \omega$.

\section{The three-body force for tetraquarks}
If $F^a$ is the colour charge operator of a quark,
for an antiquark we must have
\begin{equation}\label{FQBAR}
{\overline F}^a = -\frac{1}{2} \lambda^{a*}
\end{equation}
in order that ${\overline F}^a$ ($a$ = 1,2,...,8) satisfy the Lie algebra too.
Then one can write the three-body interaction acting in
the  $q^2 {\overline q}$ subsystem as
\begin{equation}\label{2QQBAR}
{\overline {\mathcal C}}_{ijk} = -d^{abc} ~F^{a}_i
~F^{b}_j ~ {\overline F}^{c}_k
\end{equation}
and the three-body interaction acting in
the  $q {\overline q}^2$ subsystem as
\begin{equation}\label{Q2QBAR}
{\overline {\mathcal C}}_{ijk} = d^{abc} ~F^{a}_i
~{\overline F}^{b}_j ~ {\overline F}^{c}_k
\end{equation}
As a remark, the three-body interaction
in an antibaryon should be
\begin{equation}
{\overline {\mathcal C}}_{ijk} = - d^{abc} ~{\overline F}^{a}_i
~{\overline F}^{b}_j ~ {\overline F}^{c}_k~.
\end{equation}
In Ref. \cite{DMI} the operator (\ref{2QQBAR}) is given in terms of
SU(3) invariants as
\begin{equation}\label{EQ25}
{\overline {\mathcal C}}_{ijk} =
-\frac{1}{6} [ C^{(3)}_{i+j+ \overline k} - \frac{5}{2} C^{(2)}_{i+j}
+ \frac{50}{9} ]
\end{equation}
where $C^{(3)}_{i+j+ \overline k}$ acts on the  $q^2 {\overline q}$ subsystem
but the Casimir operator
acts only on the subsystem of $i+j$ quarks
(see Appendix A). If the quark subsystem is
in a symmetric state it gives rise to a
$q^2 {\overline q}$ $[211]_C$ state called $s$
and if it is in an antisymmetric state to a $[211]_C$ state called $a$.
Both these states have ($\lambda \mu$) = (10), which
according to (\ref{AUTOG}) gives
$\langle C^{(3)}_{i+j+ \overline k} \rangle$ = $\frac{10}{9}$.
But for the subsystem
of the $i+j$ quarks only, the SU(3) representations are different.
One has ($\lambda \mu$) = (20) for the $s$ state
and ($\lambda \mu$) = (01) for the $a$ state.  Then
the expectation value of the operator (\ref{EQ25}) is
$-\frac{5}{18}$ for $s$ and $\frac{5}{9}$ for $a$,
consistent with Table II of \cite{DMI}.

The operator (\ref{Q2QBAR}) acting on a  $q {\overline q}^2$ subsystem
can be brought to a form similar to
(\ref{EQ25}). This is
\begin{equation}\label{EQ25P}
{\overline {\mathcal C}}_{ijk} =
\frac{1}{6} [ C^{(3)}_{i+ \overline j + \overline k}
+ \frac{5}{2} C^{(2)}_{\overline j + \overline k}
- \frac{50}{9} ]
\end{equation}
The difference with respect to a $q^2 {\overline q}$ subsystem
is now that one has to
calculate the expectation value of
$C^{(3)}_{i+ \overline j + \overline k}$
for a $[221]_C$ colour
state for which ($\lambda \mu$) = (01) so that
one has now $\langle C^{(3)}_{i+ \overline j + \overline k} \rangle$
= - $\frac{10}{9}$. The subsystem of antiquarks gives for $C^{(2)}$
the same value as that for the quarks so that the
operators (\ref{2QQBAR}) and (\ref{Q2QBAR}) have the same expectation
value which leads to Eqs. (26) and (27) of Ref. \cite{DMI}
\begin{equation}
V_s= \frac{5}{18}~c~ m \omega^2~(r^{2}_{12}+r^{2}_{13}+r^{2}_{14}
+r^{2}_{23}+r^{2}_{24}+r^{2}_{34})
\end{equation}
\begin{equation}
V_a=- \frac{5}{9}~c~ m \omega^2~(r^{2}_{12}+r^{2}_{13}+r^{2}_{14}
+r^{2}_{23}+r^{2}_{24}+r^{2}_{34})
\end{equation}
One can see that the contribution of the fourth particle
is also included in these equation.
To understand this one can for example add another antiquark to
$q^2 \overline q$. This leads to the singlet colour state [222]
appearing from the direct product [211] $\times$ [11]. By construction
this singlet has an intermediate coupling both between quarks
and antiquarks. The two quarks couple either to a $\overline 3$ or
a 6 state and the antiquarks to 3 or a $\overline 6$ state.
If the particles 1 and 2 are quarks and 3 and 4 are antiquarks,
$V_a$ and  $V_s$ are the three-body contribution
to the colourless states denoted
by $ |{\overline 3}_{12} 3_{34} > $ and $|6_{12} {\overline 6}_{34} >$
respectively.

For a negative $c$, as required
for baryons described by a pure constituent quark model
(no gluon components in the wave function), the mass of
the  $|6_{12} {\overline 6}_{34} >$
state is reduced
and the mass of $ |{\overline 3}_{12} 3_{34} > $ is enhanced by a three-body
force. The situation is opposite for a positive $c$. In \cite{DMI}, $V_s$
was associated to the unobserved
sextet-sextet state, which would mean that a
positive $c$ is preferable.
The conflict can be solved by noting that the
relevant states in the present
problem are in fact linear combinations of
$ |{\overline 3}_{12} 3_{34} > $ and $|6_{12} {\overline 6}_{34} >$.
Such states are important asymptotically and they are
defined by the transformations (see e.g. Ref. \cite{BS})
\begin{equation}\label{TRANSF1}
|1_{13} 1_{24} > = \sqrt{\frac{1}{3}} |{\overline 3}_{12} 3_{34} >
+  \sqrt{\frac{2}{3}} |6_{12} {\overline 6}_{34} >~,
\end{equation}
\begin{equation}\label{TRANSF2}
|8_{13} 8_{24} > = -\sqrt{\frac{2}{3}} |{\overline 3}_{12} 3_{34} >
+  \sqrt{\frac{1}{3}} |6_{12} {\overline 6}_{34} >~.
\end{equation}
In these states the intermediate  coupling in the colour space takes
place between a quark $q$ and an antiquark ${\overline q}$. This gives
colour singlet $q {\overline q}$ pairs in Eq. (\ref{TRANSF1})
and colour octet ones in Eq. (\ref{TRANSF2}). Asymptotically
the energy of $|8_{13} 8_{24} >$ must become large, as such a state is not
expected to be seen.
Using the transformations (\ref{TRANSF1}) and (\ref{TRANSF2})
one obtains the contribution
of the three-body interaction in a $q^2 {\overline q}^2$ system as
\begin{equation}
\langle 1_{13} 1_{24} | {\overline {\mathcal C}}_{123} |  1_{13} 1_{24}
\rangle \propto
[\frac{1}{3} ( - \frac{5}{9}) + \frac{2}{3}~ \frac{5}{18}]~c = 0
\end{equation}
and
\begin{equation}
\langle 8_{13} 8_{24} | {\overline {\mathcal C}}_{123} |  8_{13} 8_{24}
\rangle \propto
[\frac{2}{3} ( - \frac{5}{9}) + \frac{1}{3}~ \frac{5}{18}]~c
=  - \frac{5}{18}~c
\end{equation}
which shows that with a negative $c$ one raises the
expectation value of the octet-octet above the  singlet-singlet state,
more than with $c = 0$.
This implies that the coupling between octet-octet and singlet-singlet states
due to a hyperfine splitting will be diminished, which amounts to
make a  ground state tetraquark less stable.
This seems to be
consistent with the experimental observation that no
stable tetraquark system has been seen so far.
\section{The NN interaction}
The short-range NN interaction can be studied as a $q^6$ problem.
First we give a simplified discussion by considering that the
six quarks are in a totally symmetric orbital state $[6]_O$. In such a case
the spin-flavour part of the wave function has  a $[33]_{FS}$
symmetry which combined with the colour symmetry $[222]_C$
state leads to a totally antisymmetric state.
The latter is a superposition of five colour components
given by the five following Young tableaux:
\newline
\begin{equation}
{\psi }_{1} = \renewcommand{\arraystretch}{0.5}
\begin{array}{c} $\fbox{1}\fbox{4}$ \\
$\fbox{2}\fbox{5}$ \\
$\fbox{3}\fbox{6}$ \end{array}, \hspace{4mm}
{\psi }_{2} =
\begin{array}{c} $\fbox{1}\fbox{3}$ \\
$\fbox{2}\fbox{5}$ \\
$\fbox{4}\fbox{6}$ \end{array}, \hspace{4mm}
{\psi }_{3} =
\begin{array}{c} $\fbox{1}\fbox{3}$ \\
$\fbox{2}\fbox{4}$ \\
$\fbox{5}\fbox{6}$ \end{array}, \hspace{4mm}
{\psi }_{4} =
\begin{array}{c} $\fbox{1}\fbox{2}$ \\
$\fbox{3}\fbox{5}$ \\
$\fbox{4}\fbox{6}$ \end{array}, \hspace{4mm}
{\psi }_{5} =
\begin{array}{c} $\fbox{1}\fbox{2}$ \\
$\fbox{3}\fbox{4}$ \\
$\fbox{5}\fbox{6}$ \end{array}
\label{Young}
\end{equation}
\newline
Below we give some details of our calculations of the three-body
matrix elements for a $6q$ system.  In a state
of orbital symmetry $[6]_O$, i.e. of configuration $s^6$,
all orbital matrix elements are equal so one has only to
calculate the colour matrix element:
\begin{equation}\label{TOTAL}
\frac{1}{5}\sum_{i<j<k}^6 \sum_{a,b,c} \sum_{n=1}^{5}
<\psi_n | d^{abc} F^{a}_{i} F^{b}_{j} F^{c}_{k} | \psi_n >
\end{equation}
where the factor $1/5$ comes from the normalization of the total wave
function (see e.g. \cite{FS} , chapter 10). In Appendix B
we have explicitly proved that:
\begin{equation}
\sum_{a,b,c} <\psi_2 | d^{abc} F^{a}_{1} F^{b}_{2} F^{c}_{3} | \psi_2 >
= -5/36
\label{triplet}
\end{equation}
We get the same result for $\psi_3, \psi_4$ and $\psi_5$ but not for $\psi_1$,
for which one has:
\begin{equation}
\sum_{a,b,c} <\psi_1 | d^{abc} F^{a}_{1} F^{b}_{2} F^{c}_{3} | \psi_1 >
= 10/9
\end{equation}
i.e., the result for the singlet (123), as expected.
As shown in Table 1, the matrix elements
$<\psi_n | d^{abc} F^{a}_{i} F^{b}_{j} F^{c}_{k} | \psi_n >$ differ
for different $(ijk)$ but the sum over the five states doesn't depend on the
choice of $(ijk)$; the table has been calculated for a given value of the
colour-indices $(abc)=(146)$ but the conclusion is also true for the other
values of the colour indices. Then the calculation of the matrix element
(\ref{TOTAL})
reduces to the calculation of the matrix element
of the three-quarks (123) which has to be multiplied by $C^{3}_{6}$ = 20.

Thus we obtain
\begin{equation}
\begin{array}{r}
\displaystyle
\frac{1}{5}\sum_{i<j<k}^{6} \sum_{a,b,c} \sum_{n=1}^{5}
<\psi_n | d^{abc} F^{a}_{i} F^{b}_{j} F^{c}_{k} | \psi_n > =
\frac{20}{5} \sum_{a,b,c} \sum_{n=1}^{5}
<\psi_n | d^{abc} F^{a}_{1} F^{b}_{2} F^{c}_{3} | \psi_n > \\

\\

\displaystyle = 4 (\sum_{a,b,c}
<\psi_1 | d^{abc} F^{a}_{1} F^{b}_{2} F^{c}_{3} | \psi_1 > + 4
\sum_{a,b,c} <\psi_2 | d^{abc} F^{a}_{1} F^{b}_{2} F^{c}_{3} | \psi_2 > ) \\

\\

= 4 (10/9 -20/36) = 20/9
\end{array}
\end{equation}
Therefore we can see that in the case of a three-body confining force
the value of the matrix element of a six-quark system
in the symmetry state $[6]_O$ is equal to two
times the value for a single baryon.
This situation is similar to the two-body force where the expectation
value of a 2-body operator
$ V_{2b} \propto \sum\limits_{i<j}F^{a}_{i} F^{a}_{j} $
in the symmetry state $[6]_O$ is equal to - 4 i.e. two times
the value of a single baryon.
Let us define an adiabatic NN potential as the difference between the
interaction Hamiltonian at zero separation distance and at infinity, i.e.
\begin{equation}
V_{NN} = H(0) - H(\infty)
\end{equation}
In this difference only the kinetic energy survives
if the NN system is in the state $[6]_O$.
The contribution of
the confinement due both to two- and three-body forces cancels out
because
\begin{equation}
V^{conf}(0) = V^{conf}(\infty) = V_{2b} + V_{3b} =
-4 + 20/9 c
\end{equation}
The two-body confinement force has been discussed for
example in Ref. \cite{SPG} where the Hamiltonian also contains
a hyperfine interaction.

However the physical NN state is a combination of three symmetry
states given by \cite{HAR}
\begin{equation}\label{PHYSNN}
|NN \rangle = \sqrt{ \frac{1}{9} }~ | [6]_O [33]_{FS} >
+ \sqrt{ \frac{4}{9} }~ | [42]_O [33]_{FS} >
- \sqrt{\frac{4}{9}}~ | [42]_O [51]_{FS} >
\end{equation}
For SI=(01) or (10), one should also consider the physical $\Delta\Delta$
state:
\begin{equation}\label{PHYSDD}
|\Delta\Delta \rangle = \sqrt{ \frac{4}{45} }~ | [6]_O [33]_{FS} >
+ \sqrt{ \frac{16}{45} }~ | [42]_O [33]_{FS} >
+ \sqrt{\frac{25}{45}}~ | [42]_O [51]_{FS} >
\end{equation}
The unphysical colour octet-octet (CC) state has the form \cite{HAR}
\begin{equation}\label{CC}
|CC \rangle = \sqrt{ \frac{4}{5} }~ | [6]_O [33]_{FS} >
- \sqrt{\frac{1}{5}}~ | [42]_O [33]_{FS} >
\end{equation}

By using 3-body fractional parentage coefficients (cfp) (given in Appendix C),
we calculated the expectation value of the 3-body potential acting on the
symmetry states $| [42]_O [33]_{FS} >$ and $| [42]_O [51]_{FS} >$. For
$| [6]_O [33]_{FS} >$, the result is straightforward as shown above.
In short, we have found the following expectation values:
\begin{eqnarray}
< [6]_O [33]_{FS} | V_{3b} | [6]_O [33]_{FS} > = \frac{20}{9} c \\
< [42]_O [33]_{FS} | V_{3b} | [42]_O [33]_{FS} > = \frac{1}{9} c \\
< [42]_O [51]_{FS} | V_{3b} | [42]_O [51]_{FS} > = \frac{1}{9} c
\end{eqnarray}
Note that the $| [42]_O [33]_{FS} >$ and $| [42]_O [51]_{FS} >$ states
have the same expectation value, consistent with the fact that
$V_{3b}$ is spin-isospin independent.
Using the transformations (\ref{PHYSNN}-\ref{CC}) from the symmetry
states to the physical states $NN$, $\Delta\Delta$ and
the hidden colour $CC$ state, we obtain the
following matrix for $ V_{3b} $ :
\begin{equation}
\renewcommand{\arraystretch}{2.5}\begin{array}{c|ccc}
 & NN & \Delta\Delta & CC \\
\hline
NN & \frac{28}{81}c &
\frac{38 \sqrt{5}}{405}c & \frac{38 \sqrt{5}}{135}c \\
\Delta\Delta & \frac{38 \sqrt{5}}{405}c &
\frac{121}{405}c & \frac{76}{135}c \\
CC &  \frac{38 \sqrt{5}}{135}c &
\frac{76}{135}c & \frac{9}{5}c
\end{array}
\end{equation}
The eigenvalues of this matrix are
$E_1 = c/9, E_2 = c/9$ and $E_3 = 20 c/9$.
This shows that the effect of the 3-body colour confinement on NN and
$\Delta\Delta$ is identical and rather small as compared to that on $CC$.
In particular for a negative value of c, the spectrum of $NN$,
$\Delta\Delta$ and $CC$ lowers and shrinks.
For a positive c, the situation is the other way
round. This means
that, for $c < 0$, $V_{3b}$ itself brings some attraction and
implies a stronger coupling of $CC$ to NN and
$\Delta\Delta$ due to a hyperfine interaction.
This will lead to a reduced hard core repulsion in the NN potential.
\section{Conclusions}
We discussed the role of a schematic three-body confinement force in
the spectra of $3q,~ q^2 {\overline q}^2$ and $q^6$ systems.
We found that a three-body confinement interaction with a negative
strength $c$ has the following effects: 1) it increases the gap between
the physical colour singlet state and the unphysical coloured states
in baryons; 2) it raises the expectation value
of $q \overline q$ pairs in colour octet-octet
states with respect to singlet-singlet states in tetraquark systems,
which will lead to a smaller binding in a ground state tetraquark when a
hyperfine interaction is included and 3) it increases the coupling between
physical states and  $CC$ states in
$q^6$ systems. While, in the first two cases, the gap between the physical
colour-singlet state and the non-physical coloured states is increased, the
opposite is true for the $6q$ system. The larger coupling between
physical and $CC$ states induced by the 3-body interaction has both negative
and positive consequences: it will reinforce the undesirable Van der Waals
forces but, on the other hand, it brings more attraction into the
NN potential. This may be a desired feature for quark
models which give a too strong hard core repulsion.

A three-body force with a negative strength $c$ will have just opposite
effects than the ones mentioned above.

Our results
are valid for any quark model, irrespective of the hyperfine
interaction.
It would be useful to extend this study to a more realistic confinement
interaction.

Note also that our conclusions for $q^2\bar{q}^2$ and $q^6$
correspond to a zero separation between the interacting clusters (2 mesons
and 2 baryons respectively). It may be possible that the contribution of
the confinement interaction changes with the separation distance. This is
the aim of a further study.
\par

\vspace{1cm}
\appendix
\section{}
In this Appendix we first prove Eq. (\ref{EQ4}). In order to avoid any
confusion, we fix
the indices $(ijk)$ of (\ref{EQ4}) to be $(123)$.
We rewrite Eq. (\ref{3B}) as
\begin{eqnarray}
d^{abc}~F^{a}_{1} F^{b}_{2} F^{c}_{3}  & = &
\frac{1}{6}~[ \sum\limits_{i,j,k}^{3} d^{abc}~F^{a}_{i} F^{b}_{j} F^{c}_{k}
\nonumber \\
& & - 3 \sum\limits_{i,j}^{3} d^{abc}~F^{a}_{i} F^{b}_{i} F^{c}_{j}
+ 2 \sum\limits_{i}^{3} d^{abc}~F^{a}_{i} F^{b}_{i} F^{c}_{i}]
\label{app1}
\end{eqnarray}
where the second sum in the right-hand side compensates for
the extra terms contained in the first sum, but as we extract too many
we add the third term for satisfying the equality correctly.
The first term is precisely the cubic invariant operator acting
on the three-quark system $C^{(3)}_{1+2+3}$ and the last term
is 2 $\times$ 3 times the cubic invariant operator $C^{(3)}_1$ acting on a
quark.
The latter is replaced by its eigenvalue $\frac{10}{9}$ so we get
\begin{equation}
d^{abc}~F^{a}_{1} F^{b}_{2} F^{c}_{3}   =
\frac{1}{6}~[ C^{(3)}_{1+2+3}
 - 3 \sum\limits_{i,j}^{3} d^{abc}~F^{a}_{i} F^{b}_{i} F^{c}_{j}
+ \frac{20}{3}]
\end{equation}
Due to the fact that the constants $d^{abc}$ are symmetric under the
permutation
of indices we can modify the second term as
\begin{eqnarray}
d^{abc}~F^{a}_{1} F^{b}_{2} F^{c}_{3}  & = &
\frac{1}{6}~[ C^{(3)}_{1+2+3} \nonumber \\
& & - \frac{3}{2} \sum\limits_{i,j}^{3} d^{abc}~\{F^{a}_{i}, F^{b}_{i}\}
 F^{c}_{j}
+ \frac{20}{3}]
\end{eqnarray}
and simplify it by using the anticommutator
(\ref{ANTIC}) in the form
\begin{equation}\label{ANTICF}
\{ F^a_i, F^b_i \} = d^{abc} F^c_i
\end{equation}
and the orthogonality relation (\ref{ORTHOG}).
This leads to
\begin{equation}\label{EQ4P}
d^{abc}~F^{a}_{1} ~F^{b}_{2}~ F^{c}_{3} =
\frac{1}{6}~ [~  C^{(3)}_{1+2+3} - \frac{5}{2} C^{(2)}_{1+2+3} +
\frac{20}{3}~ ]
\end{equation}
i.e. Eq. (\ref{EQ4}).

Next we prove Eq. (\ref{EQ25}). One can rewrite the operator (\ref{2QQBAR})
as
\begin{eqnarray}
- d^{abc}~F^{a}_{1} F^{b}_{2} \overline F^{c}_{3}  & = &
-\frac{1}{6}~[ d^{abc}~(F^{a}_{1}+F^{a}_{2}+
\overline F^{a}_{3}) (F^{b}_{1}+F^{b}_{2}+
\overline F^{b}_{3}) (F^{c}_{1}+F^{c}_{2}+
\overline F^{c}_{3})
\nonumber \\
& & - 3 \sum\limits_{i,j}^{2} d^{abc}~F^{a}_{i} F^{b}_{i} F^{c}_{j}
+ 2 \sum\limits_{i}^{2} d^{abc}~F^{a}_{i} F^{b}_{i} F^{c}_{i}
\nonumber \\
& &  - 3 \sum\limits_{i}^{2} d^{abc}~F^{a}_{i} F^{b}_{i}
\overline F^{c}_{3}
\nonumber \\
& &  - 3 \sum\limits_{i}^{2} d^{abc}~F^{a}_{i} \overline F^{b}_{3}
\overline F^{c}_{3}
\nonumber \\
& &- d^{abc}~\overline F^{a}_{3} \overline F^{b}_{3}
\overline F^{c}_{3}]
\end{eqnarray}
The operator in the first term of the right-hand side is the $C^{(3)}$
invariant associated with the
whole system formed of the quarks $1$ and $2$ and the antiquark
$\overline 3$. As in Eq. (\ref{app1}), the extra-terms introduced by this
operator must be compensated in order to recover the left-hand side. This
is the role of the other terms.

The first term can be replaced by $C^{(3)}_{1+2+ \overline 3}$
and the second by 5/2~$C^{(2)}_{1+2}$ where the factor 5/2 has the same
explanation as in Eq. (\ref{EQ4P}). The third term is 2 $\times$ 2 the
invariant operator for a single quark  $C^{(3)}_{1}$ = 10/9.
The last term contains
only antiquark charge operators and is thus identical to
$C^{(3)}_{\overline 3}$
= - 10/9.
The sum over the constant terms gives
\begin{equation}
4C^{(3)}_{q} - C^{(3)}_{\overline q} = 50/9
\end{equation}
where $q = 1$ or 2 and $\overline q = \overline 3$,
so we have
\begin{eqnarray}
- d^{abc}~F^{a}_{1} F^{b}_{2} \overline F^{c}_{3}  & = &
- \frac{1}{6}~ [ C^{(3)}_{1+2+ \overline 3} - \frac{5}{2} C^{(2)}_{1+2}
- 3 \sum\limits_{i}^{2} d^{abc}~F^{a}_{i} F^{b}_{i}
\overline F^{c}_{3}
- 3 \sum\limits_{i}^{2} d^{abc}~F^{a}_{i} \overline F^{b}_{3}
\overline F^{c}_{3} + \frac{50}{9}]
\end{eqnarray}
Due to the fact that $d^{abc}$ are symmetric under permutation of
indices $a, b$ and $c$ we can again use the identity
\begin{equation}
d^{abc}~F^{a}_{i} F^{b}_{i} \overline F^{c}_{3} =
\frac{1}{2}~d^{abc}~\{ F^{a}_{i}, F^{b}_{i} \}
\overline F^{c}_{3}
\end{equation}
and
\begin{equation}
d^{abc}~F^{a}_{i} \overline F^{b}_{3} \overline F^{c}_{3} =
\frac{1}{2}~d^{abc}~F^{a}_{i}\{ \overline F^{b}_{3}, \overline F^{c}_{3} \}
\end{equation}
From (\ref{ANTIC}) and (\ref{FQBAR}) it follows that
\begin{equation}\label{ANTICFBAR}
\{ \overline F^a_i,  \overline F^b_i \} = - d^{abc}~\overline F^{c}_i~.
\end{equation}
From (\ref{ANTICF}) and (\ref{ANTICFBAR})
it follows that the third and fourth term compensate each other and
we get
\begin{equation}
{\overline {\mathcal C}}_{123} =
- \frac{1}{6}~ [ C^{(3)}_{1+2+ \overline 3} - \frac{5}{2}C^{(2)}_{1+2}  +
\frac{50}{9} ]
\end{equation}
i.e. equation (\ref{EQ25}).
\section{}
In this Appendix we give the details leading to Eq. (\ref{triplet}).
We first present the explicit expressions for the colour wave functions
corresponding to the Young tableaux of Eq. (\ref{Young}). In the
state $\psi_1$, the sets of particles $(123)$ and $(456)$ are both in a
totally antisymmetric state. Therefore, one can write:
\begin{equation}
\psi_1 = {\small \frac{1}{6} \left| \begin{array}{ccc} r(1) & b(1) & g(1) \\
r(2) & b(2) & g(2) \\ r(3) & b(3) & g(3) \end{array} \right|
 \left| \begin{array}{ccc} r(4) & b(4) & g(4) \\
r(5) & b(5) & g(5) \\ r(6) & b(6) & g(6) \end{array} \right| }
\end{equation}
where $r,b,g$ denotes the different quark colours. By applying the
permutation $(34)$ to $\psi_1$, one gets \cite{FS} :
\begin{equation}
(34) \psi_1 = \frac{1}{3} \psi_1 + \frac{2 \sqrt{2}}{3} \psi_2
\end{equation}
from where one obtains :
\begin{eqnarray}
\psi_2 & = & \frac{3}{2\sqrt{2}} \left[ (34) \psi_1 - \frac{1}{3}
\psi_1 \right]
\\
   &  = & {\small \frac{1}{4\sqrt{2}}  \left( \left| \begin{array}{ccc}
r(1) & b(1) & g(1) \\
r(2) & b(2) & g(2) \\ r(4) & b(4) & g(4) \end{array} \right|
 \left| \begin{array}{ccc} r(3) & b(3) & g(3) \\
r(5) & b(5) & g(5) \\ r(6) & b(6) & g(6) \end{array} \right|
- \frac{1}{3}  \left| \begin{array}{ccc} r(1) & b(1) & g(1) \\
r(2) & b(2) & g(2) \\ r(3) & b(3) & g(3) \end{array} \right|
 \left| \begin{array}{ccc} r(4) & b(4) & g(4) \\
r(5) & b(5) & g(5) \\ r(6) & b(6) & g(6) \end{array} \right| \right)}
\label{psi2}
\end{eqnarray}
The states $\psi_3$,$\psi_4$ and $\psi_5$ can be obtained by a similar
procedure. They read:
\begin{equation}
\begin{array}{c}
\psi_3 = {\small \frac{1}{2\sqrt{6}} \left( \left| \begin{array}{ccc}
r(1) & b(1) & g(1) \\
r(2) & b(2) & g(2) \\ r(5) & b(5) & g(5) \end{array} \right|
 \left| \begin{array}{ccc} r(3) & b(3) & g(3) \\
r(4) & b(4) & g(4) \\ r(6) & b(6) & g(6) \end{array} \right|
- \frac{1}{2} \left| \begin{array}{ccc}
r(1) & b(1) & g(1) \\
r(2) & b(2) & g(2) \\ r(4) & b(4) & g(4) \end{array} \right|
 \left| \begin{array}{ccc} r(3) & b(3) & g(3) \\
r(5) & b(5) & g(5) \\ r(6) & b(6) & g(6) \end{array} \right| \right. }\\

\\
{\small \left. + \frac{1}{2} \left| \begin{array}{ccc}
r(1) & b(1) & g(1) \\
r(2) & b(2) & g(2) \\ r(3) & b(3) & g(3) \end{array} \right|
 \left| \begin{array}{ccc} r(4) & b(4) & g(4) \\
r(5) & b(5) & g(5) \\ r(6) & b(6) & g(6) \end{array} \right| \right) }
\end{array}
\end{equation}
\begin{equation}
\begin{array}{c}
\psi_4 = {\small \frac{1}{2\sqrt{6}} \left( \left| \begin{array}{ccc}
r(1) & b(1) & g(1) \\
r(3) & b(3) & g(3) \\ r(4) & b(4) & g(4) \end{array} \right|
 \left| \begin{array}{ccc} r(2) & b(2) & g(2) \\
r(5) & b(5) & g(5) \\ r(6) & b(6) & g(6) \end{array} \right|
- \frac{1}{2} \left| \begin{array}{ccc}
r(1) & b(1) & g(1) \\
r(2) & b(2) & g(2) \\ r(4) & b(4) & g(4) \end{array} \right|
 \left| \begin{array}{ccc} r(3) & b(3) & g(3) \\
r(5) & b(5) & g(5) \\ r(6) & b(6) & g(6) \end{array} \right| \right. }\\

\\
{\small \left. + \frac{1}{2} \left| \begin{array}{ccc}
r(1) & b(1) & g(1) \\
r(2) & b(2) & g(2) \\ r(3) & b(3) & g(3) \end{array} \right|
 \left| \begin{array}{ccc} r(4) & b(4) & g(4) \\
r(5) & b(5) & g(5) \\ r(6) & b(6) & g(6) \end{array} \right| \right) }
\end{array}
\end{equation}
\begin{equation}
\begin{array}{c}
\psi_5 = {\small \frac{1}{3\sqrt{2}} \left( \left| \begin{array}{ccc}
r(1) & b(1) & g(1) \\
r(3) & b(3) & g(3) \\ r(5) & b(5) & g(5) \end{array} \right|
 \left| \begin{array}{ccc} r(2) & b(2) & g(2) \\
r(4) & b(4) & g(4) \\ r(6) & b(6) & g(6) \end{array} \right|
- \frac{1}{2} \left| \begin{array}{ccc}
r(1) & b(1) & g(1) \\
r(2) & b(2) & g(2) \\ r(5) & b(5) & g(5) \end{array} \right|
 \left| \begin{array}{ccc} r(3) & b(3) & g(3) \\
r(4) & b(4) & g(4) \\ r(6) & b(6) & g(6) \end{array} \right| \right. }\\

\\
{\small \left. - \frac{1}{2} \left| \begin{array}{ccc}
r(1) & b(1) & g(1) \\
r(3) & b(3) & g(3) \\ r(4) & b(4) & g(4) \end{array} \right|
 \left| \begin{array}{ccc} r(2) & b(2) & g(2) \\
r(5) & b(5) & g(5) \\ r(6) & b(6) & g(6) \end{array} \right|
+ \frac{1}{4} \left| \begin{array}{ccc}
r(1) & b(1) & g(1) \\
r(2) & b(2) & g(2) \\ r(4) & b(4) & g(4) \end{array} \right|
 \left| \begin{array}{ccc} r(3) & b(3) & g(3) \\
r(5) & b(5) & g(5) \\ r(6) & b(6) & g(6) \end{array} \right| \right. } \\

\\
{\small \left. - \frac{3}{4} \left| \begin{array}{ccc}
r(1) & b(1) & g(1) \\
r(2) & b(2) & g(2) \\ r(3) & b(3) & g(3) \end{array} \right|
 \left| \begin{array}{ccc} r(4) & b(4) & g(4) \\
r(5) & b(5) & g(5) \\ r(6) & b(6) & g(6) \end{array} \right|
\right) }
\end{array}
\end{equation}

We want now to calculate the expression $\sum_{a,b,c}
<\psi_2 | d^{abc} F^{a}_{1} F^{b}_{2} F^{c}_{3} | \psi_2 >$. This can be
done explicitly using Eq. (\ref{psi2}).
The non vanishing terms in this sum are given in
Table \ref{col}. Taking into account the multiplicity of each term,
one can directly check that the final result is \mbox{-5/36}.
Similar calculations can be done for the functions $\psi_3$, $\psi_4$ and
$\psi_5$ leading to the same answer.
\section{}
In this appendix, we give the values of the 3-body coefficients of
fractional parentage (cfp) necessary to calculate the expectation value of
the 3-body potential and we sketch the method to determine them.

The state $|[42]_O [33]_{FS}>$ can be decomposed as:
\begin{equation}
\begin{array}{c}
|[42]_O [33]_{FS}> = \sqrt{\frac{1}{5}} \renewcommand{\arraystretch}{0.5}
{\begin{array}{c} $\fbox{1}\fbox{2}$ \\
$\fbox{3}\fbox{4}$ \\
$\fbox{5}\fbox{6}$ \end{array} }_{OC}
{\begin{array}{c} $\fbox{1}\fbox{3}\fbox{5}$ \\
$\fbox{2}\fbox{4}\fbox{6}$
 \end{array}}_{FS} -\sqrt{\frac{1}{5}} \renewcommand{\arraystretch}{0.5}
{\begin{array}{c} $\fbox{1}\fbox{3}$ \\
$\fbox{2}\fbox{4}$ \\
$\fbox{5}\fbox{6}$ \end{array} }_{OC}
{ \begin{array}{c} $\fbox{1}\fbox{2}\fbox{5}$ \\
$\fbox{3}\fbox{4}\fbox{6}$
 \end{array}}_{FS} +\sqrt{\frac{1}{5}} \renewcommand{\arraystretch}{0.5}
{\begin{array}{c} $\fbox{1}\fbox{3}$ \\
$\fbox{2}\fbox{5}$ \\
$\fbox{4}\fbox{6}$ \end{array}}_{OC}
{\begin{array}{c} $\fbox{1}\fbox{2}\fbox{4}$ \\
$\fbox{3}\fbox{5}\fbox{6}$
 \end{array}}_{FS}  \\
\\
- \sqrt{\frac{1}{5}} \renewcommand{\arraystretch}{0.5}
{\begin{array}{c} $\fbox{1}\fbox{3}$ \\
$\fbox{2}\fbox{5}$ \\
$\fbox{4}\fbox{6}$ \end{array}}_{OC}
{\begin{array}{c} $\fbox{1}\fbox{2}\fbox{4}$ \\
$\fbox{3}\fbox{5}\fbox{6}$
 \end{array}}_{FS} -
\sqrt{\frac{1}{5}} \renewcommand{\arraystretch}{0.5}
{\begin{array}{c} $\fbox{1}\fbox{4}$ \\
$\fbox{2}\fbox{5}$ \\
$\fbox{3}\fbox{6}$ \end{array}}_{OC}
{\begin{array}{c} $\fbox{1}\fbox{2}\fbox{3}$ \\
$\fbox{4}\fbox{5}\fbox{6}$
 \end{array}}_{FS}
\end{array}
\end{equation}
One has to determine the 3-body cfp
associated to the decomposition of the $OC$ part of the wave function into
its orbital and colour parts, for example:
\begin{equation}
\renewcommand{\arraystretch}{0.5}
{\begin{array}{c}
$\fbox{\vphantom{1}\hphantom{1}}\fbox{\vphantom{1}\hphantom{1}}$ \\
$\fbox{\vphantom{1}\hphantom{1}}\fbox{4}$ \\
$\fbox{5}\fbox{6}$ \end{array}}_{OC}
\rightarrow
\renewcommand{\arraystretch}{0.5}
\begin{array}{l}
$\fbox{\vphantom{1}\hphantom{1}}\fbox{\vphantom{1}\hphantom{1}}\fbox{5}\fbox{6}$ \\
$\fbox{\vphantom{1}\hphantom{1}}\fbox{4}$ \mbox{\hspace{1mm}$_O$}  \end{array}
\renewcommand{\arraystretch}{0.5}
{\begin{array}{c}
$\fbox{\vphantom{1}\hphantom{1}}\fbox{\vphantom{1}\hphantom{1}}$ \\
$\fbox{\vphantom{1}\hphantom{1}}\fbox{4}$ \\
$\fbox{5}\fbox{6}$ \end{array}}_C
\end{equation}

To determine the 3-body cfp we need to
write the Clebsch-Gordan (CG) coefficients of $S_6$ specifying the place
of the last three particles (pqr), where p,q,r represent the row in the
Young tableau where the particles 6, 5 and 4 are located.
The position of the remaining particles is denoted shortly by $y$. By using the
factorization properties of the CG \cite{FBS}, one gets the following
relations :
\begin{equation}
\begin{array}{lcl}
S([f']p'q'r'y'[f'']p''q''r''y''|[f]pqry) &=&
K([f']p'[f'']p''|[f]p) \\
 & & \times S([f'_{p'}]q'r'y'[f''_{p''}]q''r''y''|[f_p]qry) \\
 &=& K\left([f']p'[f'']p''|[f]p\right)
K\left([f'_{p'}]q'[f''_{p''}]q''|[f_p]q\right) \\
 & & \times  S\left([f'_{p'q'}]r'y'[f''_{p''q''}]r''y''|[f_{pq}]ry\right) \\
 &=& K\left([f']p'[f'']p''|[f]p\right)
K\left([f'_{p'}]q'[f''_{p''}]q''|[f_p]q\right) \\
 & & \times K\left([f'_{p'q'}]r'[f''_{p''q''}]r''|[f_{pq}]r\right) \\
 & & \times  S\left([f'_{p'q'r'}]y'[f''_{p''q''r''}]y''|[f_{pqr}]y\right)
\end{array}
\label{factor}
\end{equation}
where the quantities $K$ are isoscalar factors
and $S$ are CG coefficients.
In particular the last factor is the
CG of $S_3$. We use the same notations as in Ref.\cite{FBS} : $[f_p]$
corresponds to the partition of $S_5$ obtained after removal of the particle
6, $[f_{pq}]$ to the partition of $S_4$ obtained after removal of the particle
5, etc.

The 3-body cfp is defined as:
\begin{equation}
\begin{array}{lcl}
K_3([f']p'q'r'[f'']p''q''r''|[f]pqr) &=& K\left([f']p'[f'']p''|[f]p\right)
 K\left([f'_{p'}]q'[f''_{p''}]q''|[f_p]q\right) \\
& & \times  K\left([f'_{p'q'}]r'[f''_{p''q''}]r''|[f_{pq}]r\right)
\end{array}
\end{equation}
The values of the $K_3$ can then be calculated by using the corresponding
tables of Ref.\cite{FBS}. They are listed in Tables \ref{cg3} and \ref{cg3b}.
They give respectively the cfp relevant for the decomposition of the
$[222]_{OC}$ and $[21^4]_{OC}$ state. In the calculation of the
expectation values, the CG of $S_3$ are not
necessary, as they are added up in the orthogonality relation.

\vspace{2cm}

{\bf Acknowledgements}.
We acknowledge warm hospitality
and living support from ECT*, Trento where this collaboration has
started. We are most grateful to David Brink for stimulating discussions
at an early stage of this study and for simplifying the proof of
Eq. (\ref{EQ4}) given in Appendix A. Useful discussions with
Jean-Marc Richard and Mitja Rosina are also gratefully acknowledged.

\begin{table}
\parbox{18cm}{\caption[matrix]{\label{mat} Examples of the matrix elements
$\langle \psi_n | d^{abc} F^{a}_{i} F^{b}_{j} F^{c}_{k} | \psi_n \rangle$
for a few
values of the indices $(ijk)$
at fixed $(abc) = (146)$. The successive columns correspond to the
states $\psi_1$,...,$\psi_5$; the last column gives the sum over the five
states. \\}}
\begin{tabular}{c|c|c|c|c|c|c}
$(ijk)$ & $\psi_1$ & $\psi_2$ & $\psi_3$ & $\psi_4$ & $\psi_5$ &
$\sum_{n=1}^5$ \\
\hline
(123) & 1/48 & -1/384 & -1/384 & -1/384 & -1/384 & 1/96 \\
(145) & 0 & -1/768 & 1/768 & 1/768 & 7/768 & 1/96 \\
(124) & 0 & 7/384 & -1/384 & -1/384 & -1/384 & 1/96 \\
\end{tabular}
\end{table}
\begin{table}
\parbox{18cm}{\caption[colour]{\label{col} Values of the non-vanishing
matrix elements
$\langle \psi_2 | d^{abc} F^{a}_{1} F^{b}_{2} F^{c}_{3} | \psi_2 \rangle $.
The first column gives the colour indices $(abc)$; the
second the corresponding constant $d^{abc}$ and the third column the value
of the matrix element.\\}}
\begin{tabular}{c|c@{\hspace{35mm}}|c@{\hspace{2cm}}}
$(abc)$ & $d^{abc}$ & $\langle \psi_2 | d^{abc} F^{a}_{1} F^{b}_{2}
F^{c}_{3} | \psi_2 \rangle $  \\
\hline
118 & $\sqrt{3}/3$ & -1/288 \\
146 & 1/2 & -1/384 \\
157 & 1/2 & -1/384 \\
228 & $\sqrt{3}/3$ & -1/288 \\
247 & -1/2 & -1/384 \\
256 & 1/2 & -1/384 \\
338 & $\sqrt{3}/3$ & -1/288 \\
344 & 1/2 & -1/384 \\
355 & 1/2 & -1/384 \\
366 & -1/2 & -1/384 \\
377 & -1/2 & -1/384 \\
448 & -$\sqrt{3}/6$ & -1/1152 \\
558 & -$\sqrt{3}/6$ & -1/1152 \\
668 & -$\sqrt{3}/6$ & -1/1152 \\
778 & -$\sqrt{3}/6$ & -1/1152 \\
888 & -$\sqrt{3}/3$ & -1/288 \\
\end{tabular}
\end{table}
\begin{table}
\caption[cg]{\label{cg3} The 3-body cfp
$K_3([42]p'q'r' [222] p''q''r''| [222] pqr)$. The rows correspond to $p'q'r'$
and the columns to $p''q''r''$. The value of $pqr$ is given in the upper-left
corner of the table.\\}
\renewcommand{\baselinestretch}{1.5}
\small\normalsize
\vspace{5mm}
\begin{tabular}{c|ccc}
$pqr=332$ & 332 & 323 & 321 \\
\hline
 221 & -$\sqrt{2/12}$ & &  \\
 212 & & $\sqrt{10/108}$ &  \\
 211 & & -$\sqrt{10/108}$ & $\sqrt{10/108}$ \\
 122 & & -$\sqrt{5/108}$ & \\
 121 & & $\sqrt{5/108}$ & -$\sqrt{5/108}$ \\
 112 & -$\sqrt{5/12}$ & &
\end{tabular}
\vspace{5mm}
\begin{tabular}{c|ccc}
$pqr=323$ & 332 & 323 & 321 \\
\hline
 221 & & $\sqrt{4/54}$  &  \\
 212 & $\sqrt{10/108}$ & -$\sqrt{5/162}$ & \\
 211 & -$\sqrt{10/108}$ & -$\sqrt{20/162}$ & -$\sqrt{5/162}$ \\
 122 & -$\sqrt{5/108}$ & -$\sqrt{10/162}$ & \\
 121 & $\sqrt{5/108}$ & -$\sqrt{40/162}$ & -$\sqrt{10/162}$ \\
 112 & & -$\sqrt{5/108}$ & $\sqrt{5/108}$
\end{tabular}

\vspace{5mm}
\begin{tabular}{c|ccc}
$pqr=321$ & 332 & 323 & 321 \\
\hline
 221 & & & $\sqrt{4/54}$  \\
 212 & & & $\sqrt{20/162}$   \\
 211 & -$\sqrt{20/108}$ & $\sqrt{10/162}$  & \\
 122 & & & $\sqrt{40/162}$  \\
 121 & $\sqrt{5/54}$ & $\sqrt{20/162}$ & \\
 112 & & -$\sqrt{5/54}$ &
\end{tabular}
\end{table}
\begin{table}
\renewcommand{\baselinestretch}{1.5}
\small\normalsize
\caption[cg]{\label{cg3b} Same as Table \ref{cg3} but for
$K_3([42]p'q'r' [222] p''q''r''| [21^4] pqr)$.\\}
\vspace{5mm}
\begin{tabular}{c|ccc}
$pqr=543$ & 332 & 323 & 321 \\
\hline
 221 & & $\sqrt{5/27}$ & \\
 212 & -$\sqrt{4/27}$ & -$\sqrt{1/81}$ &  \\
 211 & $\sqrt{4/27}$ & -$\sqrt{4/81}$ & -$\sqrt{1/81}$ \\
 122 & $\sqrt{2/27}$ & -$\sqrt{2/81}$ & \\
 121 & -$\sqrt{2/27}$ & -$\sqrt{8/81}$ & -$\sqrt{2/81}$ \\
 112 &  & $\sqrt{2/27}$ & -$\sqrt{2/27}$
\end{tabular}

\vspace{5mm}
\begin{tabular}{c|ccc}
$pqr=541$ & 332 & 323 & 321 \\
\hline
 221 & & & $\sqrt{5/27}$  \\
 212 & & & $\sqrt{4/81}$ \\
 211 & $\sqrt{8/27}$ & $\sqrt{2/81}$ & \\
 122 & & & $\sqrt{8/81}$  \\
 121 & -$\sqrt{4/27}$ & $\sqrt{4/81}$ & \\
 112 &  & $\sqrt{4/27}$ &
\end{tabular}

\vspace{5mm}
\begin{tabular}{c|ccc}
$pqr=154$ & 332 & 323 & 321 \\
\hline
 122 & & & $\sqrt{1/5}$  \\
 121 & & -$\sqrt{2/5}$ & \\
 112 & $\sqrt{2/5}$ & &
\end{tabular}

\vspace{5mm}
\begin{tabular}{c|ccc}
$pqr=514$ & 332 & 323 & 321 \\
\hline
 212 & & & -$\sqrt{5/27}$ \\
 211 & & $\sqrt{10/27}$ &  \\
 122 & & & -$\sqrt{8/135}$  \\
 121 & & $\sqrt{16/135}$ & \\
 112 & $\sqrt{4/15}$ & &
\end{tabular}
\end{table}
\newpage
\end{document}